\documentclass[9pt,twocolumn,twoside]{osajnl}
\usepackage[justification=centering]{caption}
\usepackage{float}
\usepackage{flushend}
\journal{josaa} 

\setboolean{shortarticle}{false}

\title{Physical meaning of the deviation scale under arbitrary turbulence strengths of optical orbital angular momentum}

\author[1,2,3]{Zhiwei Tao}
\author[2,4]{Yichong Ren}
\author[2]{Azezigul Abdukirim}
\author[1,2]{Shiwei Liu}
\author[2]{Ruizhong Rao}

\affil[1]{School of Environmental Science and Optoelectronic Technology, University of
Science and Technology of China, Hefei 230026, China}
\affil[2]{Key Laboratory of Atmospheric Optics, Anhui Institute of Optics and Fine
Mechanics, Chinese Academy of Sciences, Hefei 230031, China}
\affil[3]{tzw14789@mail.ustc.edu.cn}
\affil[4]{rych@aiofm.ac.cn}

\doi{\url{http://dx.doi.org/10.1364/JOSAA.XX.XXXXXX}}

\begin{abstract}
The recently so-called deviation scale [C. M. Mabena et al., Phys. Rev. A 99,
013828 (2019)] bridges the connection between the result of the infinitesimal
propagation equation (IPE) prediction and that of the single phase screen
(SPS) approximation. Thanks to the multiple phase screen (MPS) approach, in
this paper we elaborate the physical meaning of the deviation scale: the
spatial accumulation of slight intensity modulation of incident orbital
angular momentum (OAM) carrying beam splits the original vortex into multiple
individual vortices with a topological charge (TC) of $+1$ and re-generates
the vortex-antivortex pairs with a TC of $+1$ and with a TC of $-1$, leading
to a significant deviation between these two different results only when the
disruption of this compound effect on the phase distribution of the incident
OAM-carrying beam becomes more significant. Other than that, we also show that
the appearance of the deviation scale cannot be predicted only by the Rytov variance, which can be predicted through the vortex-splitting ratio
of the received optical field alone or with the help of the normalized
propagation distance.
\end{abstract}

\setboolean{displaycopyright}{true}

\begin{document}

\maketitle

\section{INTRODUCTION}

The property that photon carries orbital angular momentum (OAM) was firstly
proven by Allen in 1992\cite{c1}, which offers a feasible way to encode
information in an infinitely large Hilbert space. It had demonstrated an
impressive significance in many applications such as modal diversity\cite{c2}%
, optical trapping\cite{c3,c4} and imaging\cite{c5,c6}, remote sensing\cite%
{c7}, and quantum information\cite{c8,c9,c10}. Unfortunately, such
advantages significantly degrade in atmospheric propagation, partly because
the helical phase $e^{il\phi}$ of an OAM-carrying beam, with $l$ the quantum
number indicating the amount of OAM $l\hbar$ carried by each photon in a
beam, becomes extremely fragile while encountering atmospheric turbulence.

Recently, most previous studies, including theories\cite{c11,c12,c13,c14,c15}
and experiments\cite{c16,c17,c18,c19}, have been devoted to investigate the
effects of atmospheric turbulence on an OAM-carrying beam using phase
perturbation approximation, i.e., single phase screen (SPS) approximation (we
call this as the SPS approximation in general referring to the Paterson C
model proposed in Ref. \cite{c11}). This model indicates that the measured OAM
expectation value in the output plane is the same as that of the input mode.
Until 2017 Lavery et al. found that this result does not hold for long
free-space link lengths or more turbulent links in terms of the experiment of
transmitting high-dimensional structured optical fields in an urban
environment\cite{c20,c21}. More importantly, in 2019 Mabena et al. gives a
precise prediction for describing the crosstalk evolution of atmospheric
propagation under arbitrary scintillation conditions\cite{c22} using the
infinitesimal propagation equation (IPE) of Roux\cite{c23,c24,c25,c26}.

However, although a vast amount of attentions have been paid to study the
crosstalk evolution of an OAM-carrying beam propagating through a turbulent
channel over past decades, \textbf{no contributions have yet elaborated the
meaning of the critical point, known as the so-called deviation scale\cite%
{c22}, which bridges the connection between the result of the SPS
approximation and that of the IPE prediction.} It is the topic of this
paper. In order to compare with Roux's previous work, in this paper we first
employ the multiple phase screen (MPS) approach to re-examine the
correctness of the minimal set of parameters, in other words, whether these
parameters remain valid to completely determine the crosstalk evolution
considering the radial-mode scrambling. Based on this re-examination, we
believe that the results of numerical simulation are equivalent to that of
the IPE prediction. Secondly, we elucidate the physical meaning of the
deviation scale through calculating the vortex-splitting ratio of the
received optical field.

The SPS approximation typically follows the assumption that scintillation is
weak enough to neglect the diffraction effect in the atmospheric propagation%
\cite{c27}. This approximation can be considered as the result of
geometrical optics. Ironically, we found that the physical meaning of the
deviation scale can be elaborated clearly when the assumption is removed and
intensity fluctuations are considered, which can be accomplished by using
so-called split-step beam propagation method\cite{c28} that is valid in all
scintillation conditions. Besides, we also reveal that the deviation begins
to appear only when the vortex-splitting ratio is beyond a specific
threshold. The interrogation of the nature of the deviation scale may be
advantageous for us to determine how and when we can employ a suitable
approximation or method to investigate the effects of atmospheric turbulence
on an OAM-carrying beam.

A qualitative explanation can be achieved from the difference curves between
the results of the IPE prediction and that of the SPS approximation. When
the Rytov variance is less than unity and the normalized propagation
distance is beneath a specific threshold, no difference is found between
these two results. Conversely, the deviation starts to appear when the
distance exceeds this threshold. From the quantitative prospective, \textbf{%
the spatial accumulation of the intensity modulation of an incident
OAM-carrying beam becomes more severe with the increasing propagation
distance, which splits the original vortex into different individual
vortices with a topological charge (TC) of }$+1$\textbf{\cite{c29} and
re-generates vortex-antivortex pairs with a TC of }$+1$\textbf{\ and a TC of 
}$-1$\textbf{\cite{c34,c35,c36}, leading to the appearance of deviation
scale only when the disruption of this compound effect on the phase
distribution of the incident OAM-carrying beam becomes more significant.} In
fact, more precise thresholds that determine the appearance of the deviation
scale for different azimuthal indices of an OAM-carrying beam are obtained
through quantitative analysis. Other than that, we conclude that the
appearance of deviation scale cannot be predicted only by the Rytov
variance, which can be predicted through the vortex-splitting ratio of the
received optical field alone or with the help of the normalized propagation
distance.

The overall structure of the main text is organized as follows. In Sec. \ref%
{sec:2}, we briefly introduce the theoretical model of the crosstalk
evolution of an OAM-carrying beam using so-called Laguerre-Gaussian (LG)
modes. In Sec. \ref{sec:3}, we first review several different physical
quantities defined in Roux's previous work\cite{c22,c37,c38,c39,c40}; then
provide some technical aspects of our numerical simulation; finally
reexamine whether the minimal set of parameters obtained by the IPE remains
valid to completely determine the crosstalk evolution considering the
radial-mode scrambling. Based on the results given in above section, Sec. %
\ref{sec:4} presents a qualitative and a quantitative explanation for the
nature of the deviation scale. Sec. \ref{sec:5} we conclude.

\begin{figure}[!h]
\centering
\includegraphics[width=0.8\linewidth]{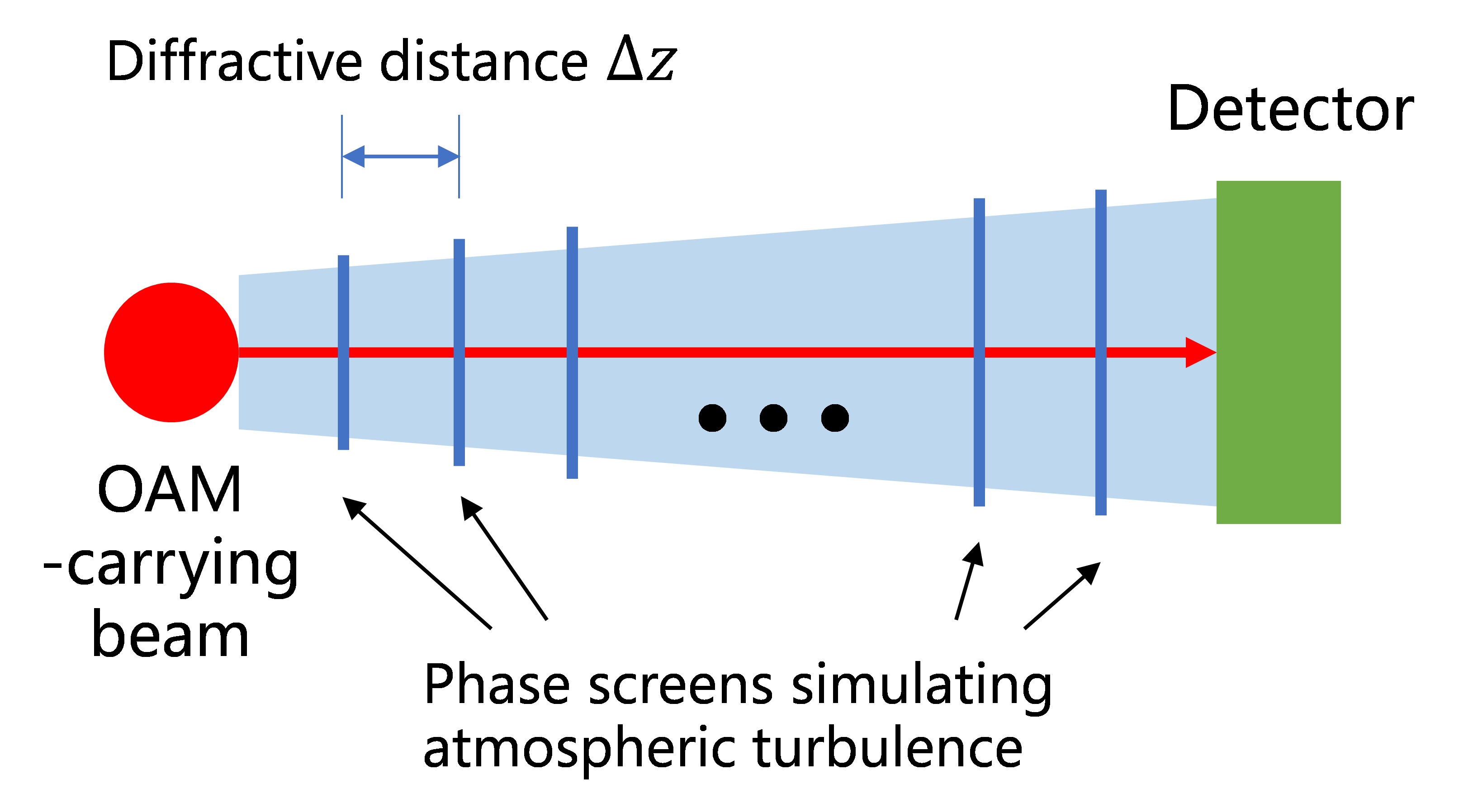}
\caption{The source generates a single LG mode and sends it through a
turbulent atmosphere (modeled by multiple turbulent cells arranged as
ladder-shaped distribution) toward a detector. The phase-screen array is
ladder-shaped distributed because the divergence angle of LG modes becomes
larger with the increase of propagation distance while holding $w_{0}$ of
the LG modes constant\protect\cite{c41}.}
\label{fig:1}
\end{figure}

\section{DESCRIPTION OF THE CROSSTALK EVOLUTION}
\label{sec:2}

The conceptual diagram of optical system that is simulated by our numerical
procedure is shown in Fig. \ref{fig:1}. Without loss of generality, we
suppose that $z$ is the propagation distance, and the incident OAM-carrying
beam is a single LG mode with azimuthal index $l_{0}$ and radial index $p_{0}
$, which can be expressed in cylindrical coordinates by

\begin{align}
LG_{p_{0},l_{0}}^{\left( q_{0}\right) }\left( r,\phi,0\right) & =A\left( 
\frac{i\sqrt{kz_{R}}r}{q_{0}}\right) ^{\left\vert l_{0}\right\vert
+1}L_{p_{0}}^{\left\vert l_{0}\right\vert }\left( \frac{kz_{R}r^{2}}{%
\left\vert q_{0}\right\vert ^{2}}\right)  \nonumber \\
& \left( -\frac{q_{0}^{\ast}}{q_{0}}\right) ^{p_{0}}\frac{e^{-\frac {ikr^{2}%
}{2q_{0}}}}{r}e^{il_{0}\phi},   \label{eq1}
\end{align}
with $L_{p_{0}}^{\left\vert l_{0}\right\vert }\left( \cdot\right) $ as the
generalized Laguerre polynomial, where $A=\sqrt{p_{0}!/\pi\left( \left\vert
l_{0}\right\vert +p_{0}\right) !}$ is the normalization constant and $%
q_{0}=iz_{R}$ represents the complex parameter associated with beam waist. $%
z_{R}=\pi w_{0}^{2}/\lambda$ and $k=2\pi/\lambda$ denote the Rayleigh range
and the wave number respectively, $\lambda$ is the wavelength and $w_{0}$ is
the beam waist at input plane.

Beam propagating through a turbulent channel is often described by the
split-step beam propagation method\cite{c28}. It generally means that, as
shown in Fig. \ref{fig:1}, a turbulent channel can be divided into a series
of turbulent cells, each cell introduces a random contribution $\phi$ to the
phase, but essentially no change in the amplitude; besides, intensity
fluctuations build up by diffraction over many cells. To agree well with the
phase structure function of the Kolmogorov turbulence model, random phase
screen lost low spatial frequencies need to be compensated using subharmonic
method\cite{c42}.

With this in mind, the split-step method is used to simulate the procedure of
atmospheric propagation instead of the SPS approximation. Since LG modes with
same $w_{0}$ form an orthonormal basis, the complex quantity $q_{0}$ is
modified to $q\left(  z\right)  =z+q_{0}$ during the spectral decomposition
for attributing the intermodal crosstalk to the impact of turbulence entirely,
where the notation $q\left(  z\right)  $ called the Siegman complex
parameter\cite{c43}. Hence, the received optical field can be decomposed into
a superposition of several LG modes%

\begin{equation}
U_{p,l}\left( r,\phi,z\right) =\sum_{p=0}^{\infty}\sum_{l=-\infty}^{\infty
}c_{p,l}LG_{p,l}^{\left( q\right) }\left( r,\phi,z\right) ,   \label{eq2}
\end{equation}
with the coefficient $c_{p,l}$ calculating from the overlap integral

\begin{equation}
c_{p,l}=\int_{0}^{\infty}rdr\int_{0}^{2\pi}d\phi U_{p,l}\left( r,\phi
,z\right) LG_{p,l}^{\left( q\right) \ast}\left( r,\phi,z\right) , 
\label{eq3}
\end{equation}
where the asterisk denotes the complex conjugate. Besides, the fraction of
optical power that is transferred from the input mode with azimuthal index $%
l_{0}$ to other modes with azimuthal index $l$ can be described as

\begin{equation}
P\left( l\right) =\sum_{p=0}^{\infty}\left\vert c_{p,l}\right\vert ^{2}, 
\label{eq4}
\end{equation}

From the above analysis, we summary the effects of atmospheric turbulence on
a LG mode with more intuitive way: the optical power of the input mode with
azimuthal index $l_{0}$ is carved into unlimited pieces in terms of the
power fraction $\left\vert c_{p,l}\right\vert ^{2}$; each piece of LG modes
attachs a turbulence-induced random phase in terms of the argument of the
overlap intergral, namely, $\arg\left( c_{p,l}\right) $.

\section{NUMERICAL REEXAMINATION OF THE IPE}
\label{sec:3}
\subsection{Several parameters}

To reexamine whether the minimal set of parameters obtained by the IPE
remains valid to completely determine the crosstalk evolution considering
the radial-mode scrambling, we briefly review several dimensionless
quantities that are theoretically indispensable in the result of the IPE
prediction: the normalized propagation distance $t\equiv z/z_{R}$, the
normalized turbulence strength $K\equiv
C_{n}^{2}w_{0}^{11/3}\pi^{3}/\lambda^{3}$ and the relative beam width $%
W\equiv w_{0}/r_{0}$, where $C_{n}^{2}$ is the refractive index structure
constant, $r_{0}$ is the Fried parameter. Besides, the compound quantities $%
W $, $K$, and $t$ are related by\cite{c22,c38,c39}

\begin{equation}
W=1.37\left( Kt\right) ^{3/5},  \label{eq5}
\end{equation}

The intensity fluctuation of an optical field propagating through atmospheric
turbulence is often centered around the scintillation index (For a more
general expression of scintillation index, we refer the reader to Ref.
\cite{c44}), which is the normalized variance of intensity fluctuation, and
can be defined as\cite{c27}%
\begin{equation}
\sigma_{I}^{2}\equiv\frac{\left\langle I^{2}\right\rangle -\left\langle
I\right\rangle ^{2}}{\left\langle I\right\rangle ^{2}}=\frac{\left\langle
I^{2}\right\rangle }{\left\langle I\right\rangle ^{2}}-1,\label{eq6}%
\end{equation}
where the notation $I$ and $\left\langle \cdot\right\rangle $ represent the
intensity of an optical field and ensemble average, respectively. Under the
weak scintillation condition based on the Kolmogorov turbulence model, the
scintillation index of a plane wave can be expressed by\cite{c27,c45}%

\begin{equation}
\sigma_{R}^{2}=1.23C_{n}^{2}k^{7/6}z^{11/6},\label{eq7}%
\end{equation}
which can also be used to describe the intensity fluctuation over the
turbulent link when extended to strong scintillation condition by increasing
either $C_{n}^{2}$ or $z$, or both. The link between these two quantities is
that under the weak scintillation condition, the scintillation index is
proportional to the Rytov variance for a plane wave. In other words, the
scintillation index increases with increasing values of the Rytov variance
until it reaches a maximum value greater than unity in the regime
characterized by random focusing\cite{c45}. Hence, we employ the Rytov
variance as an indicator of the scintillation strength to avoid the
circumstance that one scintillation index corresponds to two scintillation
conditions. The Rytov variance can be also expressed in terms of $K$ and
$t$\cite{c22,c39}%

\begin{equation}
\sigma_{R}^{2}=1.64W^{5/3}t^{5/6}=1.055W^{55/18}/K^{5/6}, \label{eq8}%
\end{equation}

\subsection{Data details and notes}

The result of numerical simulation in one iteration, representing a single
realization of turbulence, can be repeated several times. The ensemble
average of different iterations of the two steps represents the result of
crosstalk evolution under one fixed compound quantity (i.e., $K$ or $t$).
Table \ref{tab:1} gives the different values of $K$ used in our simulation.
For convenience of comparison, the turbulence strength $C_{n}^{2}$, the beam
waist $w_{0}$, and the wavelength $\lambda$ are specifically assigned to
guarantee that $W$ is limited to a certain range. Therefore, sampling
intervals and propagation distances vary with the variation of $K$ in Table %
\ref{tab:1} during these simulations.

Eq. (\ref{eq8}) gives the relationship between $\sigma _{R}^{2}$ and $K$.
Hence, it can be observed that different scintillation conditions can be
acquired during the numerical simulations by adjusting this compound
quantity with fixed $W$. Further, since the spacing between two turbulent
cells varies smartly with these parameters, different numbers of LG modes
used for spectral decomposition are set in these simulations. Finally, it is
worth highlighting that the total propagation distance ranges from a few
meters to hundred kilometers.

\subsection{Evolution rule under arbitrary scintillation conditions}

\begin{figure*}[!h]
\centering
\includegraphics[width=0.7\linewidth]{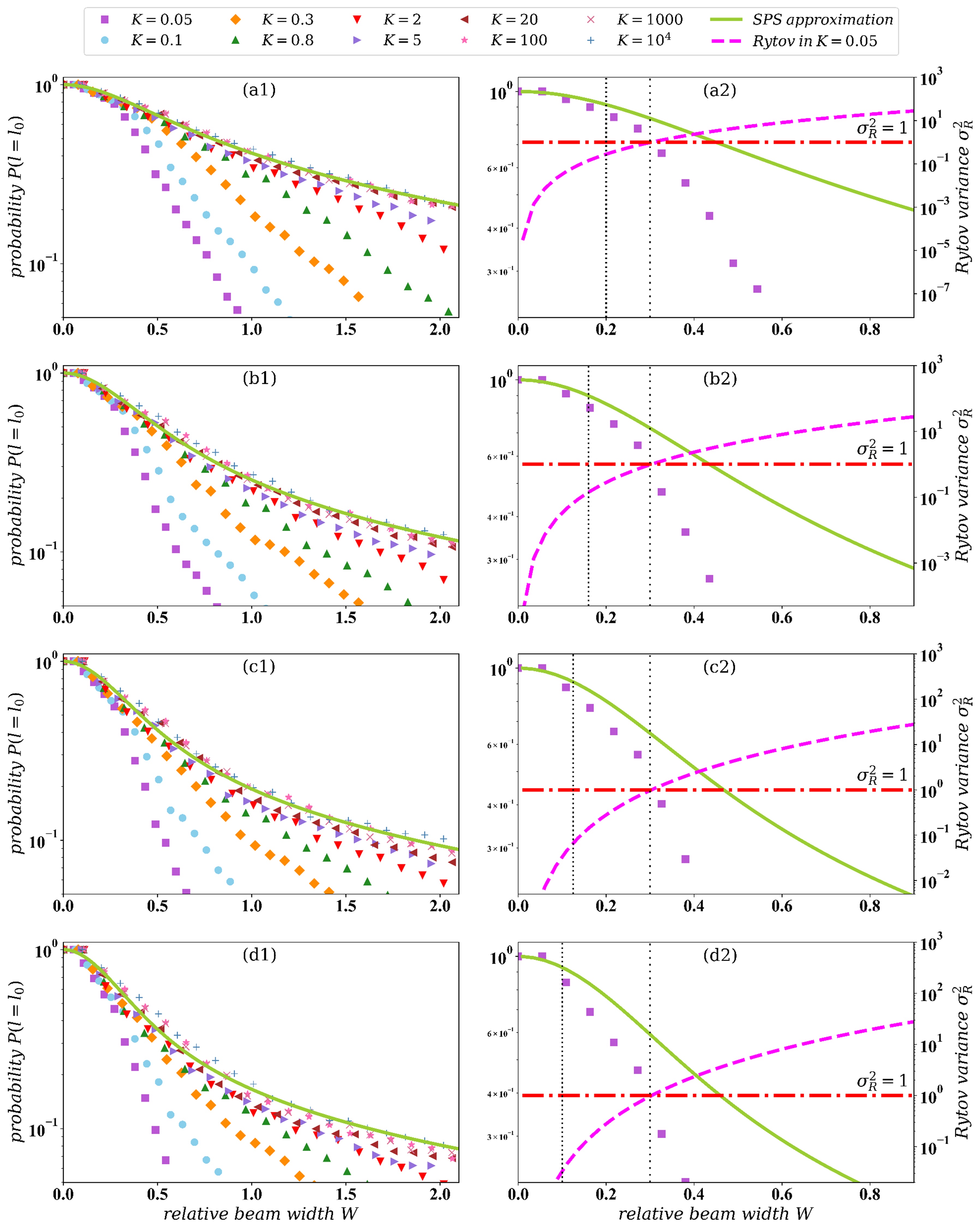}
\caption{The eigen-crosstalk probabilities averaged over 300 realizations of
turbulence as a function of $W$ under different values of normalized
turbulence strengths $K$ given in Table \protect\ref{tab:1} for (a1) $l_{0}=0
$, (b1) $l_{0}=1$, (c1) $l_{0}=2$, (d1) $l_{0}=3$. The solid lines represent
the results of the SPS approximation. The Rytov variances in $K=0.05$ (weak
turbulence) are provided in the right-column plots. The short-spacing dotted
lines and long-spacing dotted lines represent the deviation scale and the
scale determining the onset of strong scintillation, respectively. For each
azimuthal index $l_{0}$, the deviation scale appears at a smaller value
compared to the scale determining the onset of strong scintillation. The
horizontally dashed-dotted lines in all right-column plots correspond to $%
\protect\sigma _{R}^{2}=1$. The solid line and the square scatter points in
the right-column plots ((a2) $l_{0}=0$; (b2) $l_{0}=1$; (c2) $l_{0}=2$; (d2) 
$l_{0}=3$) is the partial enlargement of the results presented in the
left-column plots.}
\label{fig:2}
\end{figure*}

The fraction of optical power obtained in the output plane that remains in
the input LG mode (we call this as the eigen-crosstalk probability in this
paper; (a1) $l_{0}=0$, (b1) $l_{0}=1$, (c1) $l_{0}=2$, (d1) $l_{0}=3$) is
illustrated in Fig. \ref{fig:2}. All plots are plotted as a function of $W$.
Solid curve in all plots correspond to the results of the SPS model using
quadratic approximation\cite{c46}, while the results of our numerical
results are shown with different shaped lines, corresponding to different
values of $K$ given in Table \ref{tab:1}. According to Eq. (\ref{eq8}), the
Rytov variances in $K=0.05$ are shown in all right-column plots with dashed
lines. The dotted lines with small-spacing represent the deviation scale,
while the large-spacing ones represent the scale determining the onset of
strong scintillation. The horizontally dashed-dotted lines correspond to $%
\sigma_{R}^{2}=1$.

\begin{table}[!b]
\centering
\caption{\bf Different values of the compound quantity $K$ composed of
beam parameters and turbulence parameters shown in Fig. \ref{fig:2}.}
\begin{tabular}
[c]{cccc}%
\toprule
$K$ & $C_{n}^{2}\left(  m^{-2/3}\right)  $ & $w_{0}\left(  m\right)  $ &
$\lambda\left(  nm\right)  $\\
\midrule
$0.05$ & $10^{-16}$ & $0.05$ & $1000$\\
$0.1$ & $10^{-16}$ & $0.07$ & $1214$\\
$0.3$ & $10^{-16}$ & $0.1$ & $1314$\\
$0.8$ & $5\times10^{-16}$ & $0.05$ & $689$\\
$2$ & $5\times10^{-16}$ & $0.08$ & $900$\\
$5$ & $5\times10^{-15}$ & $0.05$ & $807$\\
$20$ & $10^{-14}$ & $0.05$ & $640$\\
$100$ & $10^{-13}$ & $0.08$ & $1432$\\
$1000$ & $10^{-12}$ & $0.08$ & $1434$\\
$10^{4}$ & $5\times10^{-12}$ & $0.1$ & $1495$\\%
\bottomrule
\end{tabular}
  \label{tab:1}
\end{table}

We observe that for large values of $K$ (e.g., $K\gtrsim 20$ for $l_{0}=0,1$
and $K\gtrsim 100$ for $l_{0}=2,3$), the evolution curves of the
eigen-crosstalk probability obtained by our numerical simulation almost
coincide with the result of the SPS approximation. However, this coincidence
does not exist as $K$ becomes smaller (e.g., $K\lesssim 5$ for $l_{0}=0,1$
and $K\lesssim 20$ for $l_{0}=2,3$), which means that another compound
quantity $K$ is needed to be added to describe the evolution of
eigen-crosstalk probability except $W$. Moreover, we found that the
eigen-crosstalk probability of input modes with a larger azimuthal index ($%
l_{0}\gtrsim 2$) decreases faster compared to that of the smaller one, more
obviously when $K\lesssim 2$, which is likely because LG modes with a larger 
$l_{0}$ possess a larger beam waist so that it's more susceptible to
atmospheric turbulence.

The deviation phenomenon between the results of numerical simulation and
that of the SPS approximation under arbitrary turbulence strengths is also
shown in Fig. \ref{fig:2}. We zoom in the results of numerical simulation of 
$K=0.05$ in the right-column of Fig. \ref{fig:2} ((a2) $l_{0}=0$, (b2) $%
l_{0}=1$, (c2) $l_{0}=2$, (d2) $l_{0}=3$). The crossing points of the
numerical curves and the large-spacing dotted lines represent the onset of
strong scintillation, while the crossing points of the numerical curves and
the short-spacing dotted lines indicate at what values of the Rytov variance
the deviation occur. Notably, the results in Fig. \ref{fig:2} seem to
indicate that LG modes with a larger azimuthal index have a smaller
deviation scale. Further, it reveals a well-known conclusion that the
deviation scale is always smaller than the scale determining the onset of
strong scintillation\cite{c22}. However, why are they different? more
straightforwardly, what is the nature of the deviation scale? It deserves to
be investigated, and it is also the topic of this paper.

\subsection{Verification of the minimal set of parameters}

\begin{figure}[!b]
\centering
\includegraphics[width=0.8\linewidth]{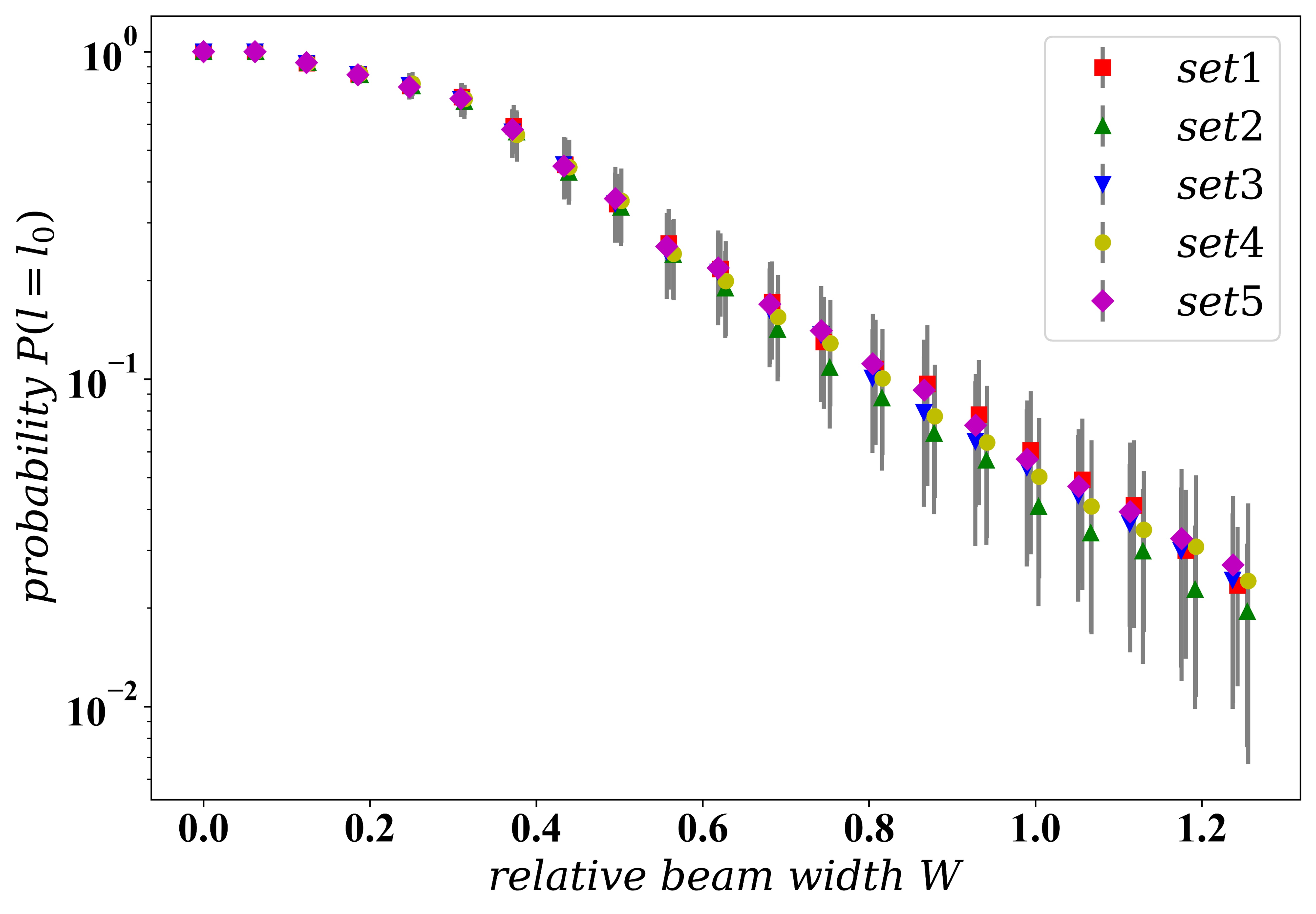}
\caption{The eigen-crosstalk probability as a function of $W$ with the same
normalized turbulence strength $K=0.066$ given in Table \protect\ref{tab:2}
for $l_{0}=0$, averaged over 300 realizations of turbulence. Each set of
dimensionless parameters are combined with different beam and turbulence
parameters. The error bars represent the standard error.}
\label{fig:3}
\end{figure}

The evolution curves of eigen-crosstalk probability as a function of $W$ for 
$l_{0}=0$ are illustrated in Fig. \ref{fig:3}. These curves are obtained in $%
K=0.066$ with different combinations of dimension parameters shown in Table %
\ref{tab:2}. The smaller $K$ is chosen because it represents a condition
where the evolution of eigen-crosstalk probability deviates significantly
from that of the SPS approximation\cite{c38}. We observe that although
different sets of parameters are selected in the numerical simulation, all
of curves coincide under the same $K$, which means all kinds of beam
parameters and turbulence parameters (i.e., $C_{n}^{2}$, $z$, $w_{0}$ and $%
\lambda $) always appear in the form of compound quantities (i.e., $K$, $t$,
and $W$) within the crosstalk evolution calculated by the Kolmogorov
turbulence model under arbitrary scintillation conditions. Compared to the
minimal set of parameters that decides the result of the IPE prediction, one
can conclude that these parameters remain valid to completely determine the
crosstalk evolution considering the radial-mode scrambling.

\begin{table}[!b]
\centering
\caption{\bf Different combinations of beam parameters and turbulence
parameters for $K=0.066$ shown in Fig. \ref{fig:3}.}
\begin{tabular}
[c]{cccc}%
\toprule
$set$ & $C_{n}^{2}\left(  m^{-2/3}\right)  $ & $w_{0}\left(  m\right)  $& $\lambda\left(  nm\right)  $\\
\midrule
$1$ & $10^{-17}$ & $0.1$ & $1000$\\
$2$ & $10^{-16}$ & $0.05$ & $924$\\
$3$ & $5\times10^{-18}$ & $0.1$ & $794$\\
$4$ & $5\times10^{-17}$ & $0.08$ & $1302$\\
$5$ & $10^{-18}$ & $0.13$ & $640$\\%
\bottomrule
\end{tabular}
  \label{tab:2}
\end{table}

\begin{table}[!h]
\centering
\caption{\bf Different values of the compound quantity $t$ composed of
beam parameters and turbulence parameters shown in Fig. \ref{fig:4}.}
\begin{tabular}
[c]{cccc}%
\toprule
$t$ & $z\left(  km\right)  $ & $w_{0}\left(  m\right)  $ & $\lambda
\left(  nm\right)  $\\
\midrule
$2\times10^{-4}$ & $5\times10^{-3}$ & $0.1$ & $1263$\\
$10^{-3}$ & $3\times10^{-2}$ & $0.08$ & $672$\\
$10^{-2}$ & $0.1$ & $0.06$ & $1169$\\
$0.1$ & $1.2$ & $0.05$ & $658$\\
$0.5$ & $5$ & $0.05$ & $791$\\
$1$ & $22.5$ & $0.08$ & $900$\\
$1.5$ & $30$ & $0.07$ & $770$\\
$2$ & $50$ & $0.08$ & $806$\\
$3$ & $70$ & $0.08$ & $864$\\
$5$ & $100$ & $0.08$ & $1000$\\
$10$ & $200$ & $0.08$ & $1000$\\%
\bottomrule
\end{tabular}
  \label{tab:3}
\end{table}

\section{PHYSICAL MEANING OF THE DEVIATION SCALE}
\label{sec:4}
\subsection{Qualitative understanding}

From above analysis, we highlight that compound quantities composed of $K$
and $W$ can completely determine the crosstalk evolution caused by
atmospheric turbulence. Consequently, it not necessarily investigates the
crosstalk evolution under different values of normalized propagation
distance $t$. However, although $K$, $t$ and $W$ are pairwise independent
compound quantities, it may be help us to give a qualitative understanding
of the deviation scale by studying the influence of $t$ on the
eigen-crosstalk probability.

\begin{figure*}[!h]
\centering
\includegraphics[width=0.7\linewidth]{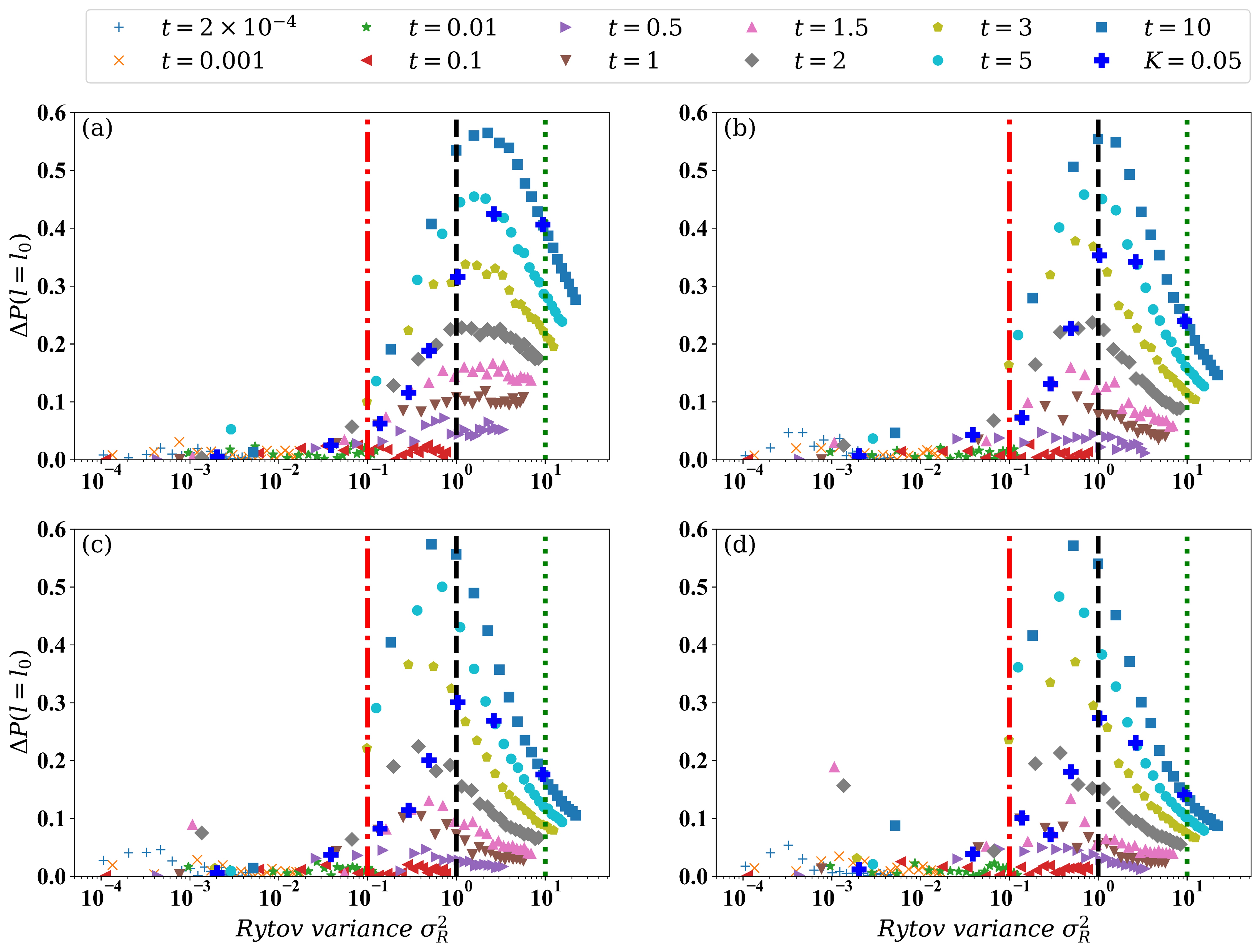}
\caption{The difference curves of eigen-crosstalk probability between the
results of the IPE prediction and that of the SPS approximation, namely, $%
\Delta P\equiv P_{SPS}-P_{IPE}$, averaged over 300 realizations of
turbulence as a function of $\protect\sigma_{R}^{2}$ under different
normalized propagation distances $t$ given in Table \protect\ref{tab:3} for
(a) $l_{0}=0$, (b) $l_{0}=1$, (c) $l_{0}=2$, (d) $l_{0}=3$. Each graph is
divided by four areas composed of $\protect\sigma_{R}^{2}<0.1$, $0.1<\protect%
\sigma_{R}^{2}<1$, $1<\protect\sigma_{R}^{2}<10$ and $\protect\sigma%
_{R}^{2}>10$. A series of cruciform scatter points are marked, representing
the results obtained in $K=0.05$.}
\label{fig:4}
\end{figure*}

To better determine the value of $\sigma _{R}^{2}$ and $t$ where the results
of the IPE prediction (we believe the results of numerical simulation are
equivalent to that of the IPE prediction due to the validity of the minimal
set of parameters) start to deviate from that of the SPS approximation, in
Fig. \ref{fig:4} we illustrate how the difference of eigen-crosstalk
probability between these two results, namely, $\Delta P\equiv
P_{SPS}-P_{IPE}$, changes with respect to $\sigma _{R}^{2}$ under different
values of $t$ shown in Table \ref{tab:3}. For convenience of analysis, Fig. %
\ref{fig:4} is divided into four areas. We see that the points where the IPE
predictions start to deviate from the SPS counterparts occur at $\sigma
_{R}^{2}\simeq 0.1$, especially for the larger value of $t$. Further, we
observe that when $0.1<\sigma _{R}^{2}<1$ and $t<0.5$, the deviation between
these two results remains still very small. On the contrary, if the above
conditions do not hold any more, the deviation will grow rapidly with the
increase of $\sigma _{R}^{2}$ even if the turbulent channel is still under
the weak scintillation condition.

We mark a series of cruciform scatter points in Fig. \ref{fig:4}, which
represents the results obtained in $K=0.05$. Based on these results, we can
observe clearly why the deviation always appears at smaller values than the
scale determining the onset of strong scintillation. That is because the
difference of eigen-crosstalk probability in $K=0.05$ have exceeded the
above threshold of $t$ in the regime of $0.1<\sigma _{R}^{2}<1.$ Actually,
this effect can be seen more clearly when Fig. \ref{fig:2} is plotted as a
function of $t$ instead of $W$.\textbf{\ }Hence, \textbf{we can
qualitatively conclude that the SPS approximation is valid if }$\sigma
_{R}^{2}<1$\textbf{\ and }$t<0.5$\textbf{\ for LG modes.}

Other than the interrogation of the nature of the deviation scale, we can
also conclude some other interesting phenomena from the results in Fig. \ref%
{fig:4}, which can be summarized as follows: When the scintillation
condition is beyond the weak range (i.e., $1<\sigma _{R}^{2}<10$), the
deviation between two results becomes smaller as the increase of $\sigma
_{R}^{2}$. When $\sigma _{R}^{2}>10$, these deviations together incline to
disappear, which is likely because of the effect of scintillation saturation.

\subsection{Quantitative understanding}

What can be seen from the above analysis is that the deviation between the
results of the IPE prediction and that of the SPS approximation comes from the
spatial accumulation of the intensity modulation of the input LG mode. In
other words, for a LG mode that are perturbed while propagating over a series
of turbulent cells the compound effect on the LG mode is more damaging than
the effect of simply accumulating the phase perturbation over the link. The
direct result of the intensity modulation changes the location of the original
vortex. Hence, we anticipate that even if the turbulent channel is under the
weak scintillation condition, the spatial accumulation of slight intensity
modulation of the input LG mode splits the vortex into multiple individual
vortices with a TC of $+1$ and re-generates the vortex-antivortex pairs with a
TC of $+1$ and with a TC of $-1$, which may be the main reason for the
appearance of the deviation scale.

\begin{figure*}[!h]
\centering
\includegraphics[width=\linewidth]{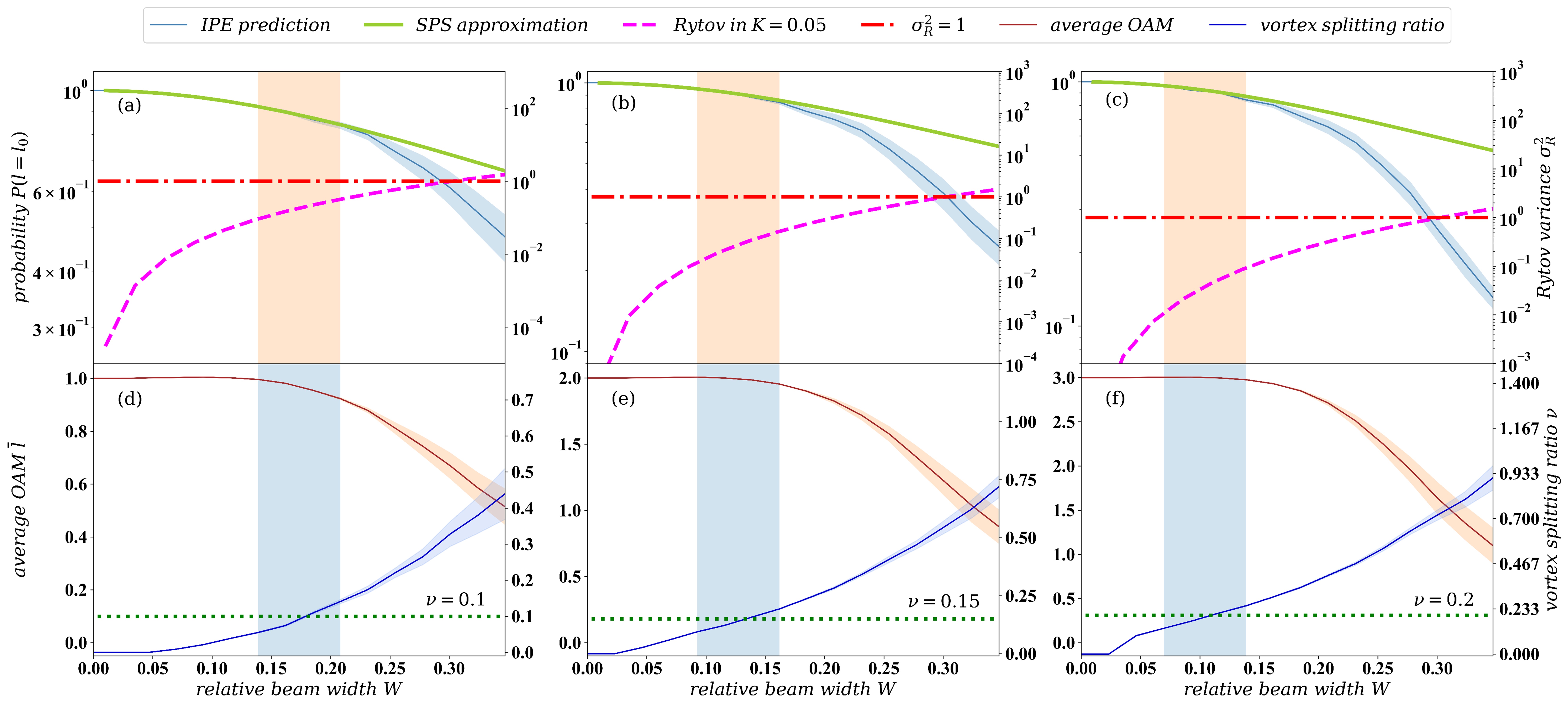}
\caption{The eigen-crosstalk probabilities ((a)--(c)), the vortex-splitting
ratio and the average OAM ((d)--(f)), averaged over 300 realizations of
turbulence as a function of $W$ with $K=0.05$: (a), (d) $l_{0}=1$, (b), (e) $%
l_{0}=2$, (c), (f) $l_{0}=3$. The standard errors are plotted by the shaded
area around each curve. The solid lines represent the results of the SPS
approximation. The Rytov variance in $K=0.05$ are depicted in dashed line.
(a)--(c) are horizontally divided into two areas composed of $\protect\sigma %
_{R}^{2}<1$ and $\protect\sigma _{R}^{2}>1$ using dashed-dotted lines. The
splitting-thresholds that represent the appearance of the deviation scale
for different azimuthal indices of LG modes are plotted by the dotted line
in (d)--(f): (d) $\protect\nu =0.1$ for $l_{0}=1$, (e) $\protect\nu =0.15$
for $l_{0}=2$, (f) $\protect\nu =0.2$ for $l_{0}=3$. These
splitting-thresholds can be easily translated into the threshold of $t$ with
different values: (d) $t=0.648$ for $l_{0}=1$, (e) $t=0.369$ for $l_{0}=2$,
(f) $t=0.255$ for $l_{0}=3$.}
\label{fig:5}
\end{figure*}

To examine our anticipation, Fig. \ref{fig:5} illustrates the evolution
curves of the eigen-crosstalk probability, the average OAM and
vortex-splitting ratio\cite{c20,c21} in the output plane, all of which are
calculated under $K=0.05$ and different values of $l_{0}$. Besides, all
plots are plotted as a function of $W$. Notably, the reason why we choose
the vortex-splitting ratio as an indicator to examine our anticipation is
that for LG modes with a specific azimuthal index, this parameter contains
much information about the vortex distribution in the phase of the received
optical field, which characterizes all the effects of atmospheric
propagation under the weak scintillation condition. Considering atmospheric
turbulence can also generate vortex-antivortex pairs, the vortex-splitting
ratio can be re-evaluated by\cite{c36}%
\begin{equation}
\nu\equiv\frac{1}{Nw_{0}}%
{\displaystyle\sum\limits_{i=1}^{N}}
q_{i}V_{r}^{\left(  i\right)  }.\label{eq9}%
\end{equation}
where $N$ represents the total number of vortices in the output plane. Each
vortex is identified by the index $i$. $q_{i}$ and $V_{r}^{\left( i\right) }$
stand for the TC of the vortex with index $i$ and the radial distance from
the beam origin for the vortex with index $i$, respectively. We have to
emphasize that this definition not only takes account into the
turbulence-induced vortex-antivortex pairs but also includes the effect of
vortex-splitting. Our vortex detection algorithm is a modified version of
the algorithm proposed in Refs. \cite{c47,c48,c49,c50}.

We observe in Fig. \ref{fig:5} that for different azimuthal indices of LG
modes and specific turbulence strength, the phenomena of vortex splitting
and vortex-antivortex pairs re-generation occur in the output plane at the
early stages of propagation (e.g., for $l_{0}=1$, atmospheric turbulence
steers the original vortex off beam axis and leads to the vortex-antivortex
pairs re-generation). Other than that, we see that the compound effect
caused by atmospheric turbulence, quantified by the vortex-splitting ratio,
becomes more pronounced progressively as $W$ increases. This behavior can be
seen more clearly when the curves are plotted as a function of $t$ instead
of $W$. Based on the results presented in Fig. \ref{fig:5}, we found that
with the increase of the vortex splitting ratio (e.g., $\nu \geq 0.1$ for $%
l_{0}=1$, $\nu \geq 0.15$ for $l_{0}=2$ and $\nu \geq 0.2$ for $l_{0}=3$),
the average OAM and the eigen-crosstalk probabilities gradually deviate from
the results of the SPS approximation. A suitable explanation is that when
the turbulent channel is under the weak scintillation condition, the
turbulence-induced intensity fluctuation splits the original vortex into
multiple vortices with a TC of $+1$ in the output plane, and possibly
accompanied by the re-generation of vortex-antivortex pairs that carry
opposite-sign unity TC. After undergoing the spatial accumulation of a
considerable amount of compound effect, the probability of detecting the TC
of $+1$ in the phase distribution of the received optical field becomes more
larger, which leads to a significant reduction of the average OAM and the deviation between the results of the IPE prediction and that of the SPS
approximation.

Therefore, we believe the appearance of the deviation is actually the
process of the spatial accumulation of the compound effect. From the results
presented in Fig. \ref{fig:5} (see the rectangular shaded areas in all six
plots), \textbf{it is worth highlighting that such phenomena will happen
only if the vortex-splitting ratio of the received optical field is beyond a
specific threshold}\textit{\ }(we call this as the splitting-threshold,
e.g., $\nu =0.1$ for $l_{0}=1$, $\nu =0.15$ for $l_{0}=2$ and $\nu =0.2$ for 
$l_{0}=3$). Other than that, \textbf{it is not difficult to find that
different splitting-thresholds that we obtained in fact provide a more
precise application scope of the SPS approximation for LG modes with a
specific azimuthal index.} To this end, we calculate the values of
normalized propagation distance $t$ that correspond to the aforementioned
splitting-thresholds by Eq. (\ref{eq5}), namely, $t=0.648$ for $l_{0}=1$, $%
t=0.369$ for $l_{0}=2$ and $t=0.255$ for $l_{0}=3$. The reason why $t$
becomes smaller as $l_{0}$ increases is that LG modes with a larger
azimuthal index possess a larger beam waist so that it's more susceptible to
atmospheric turbulence.

\begin{figure*}[!h]
\centering
\includegraphics[width=\linewidth]{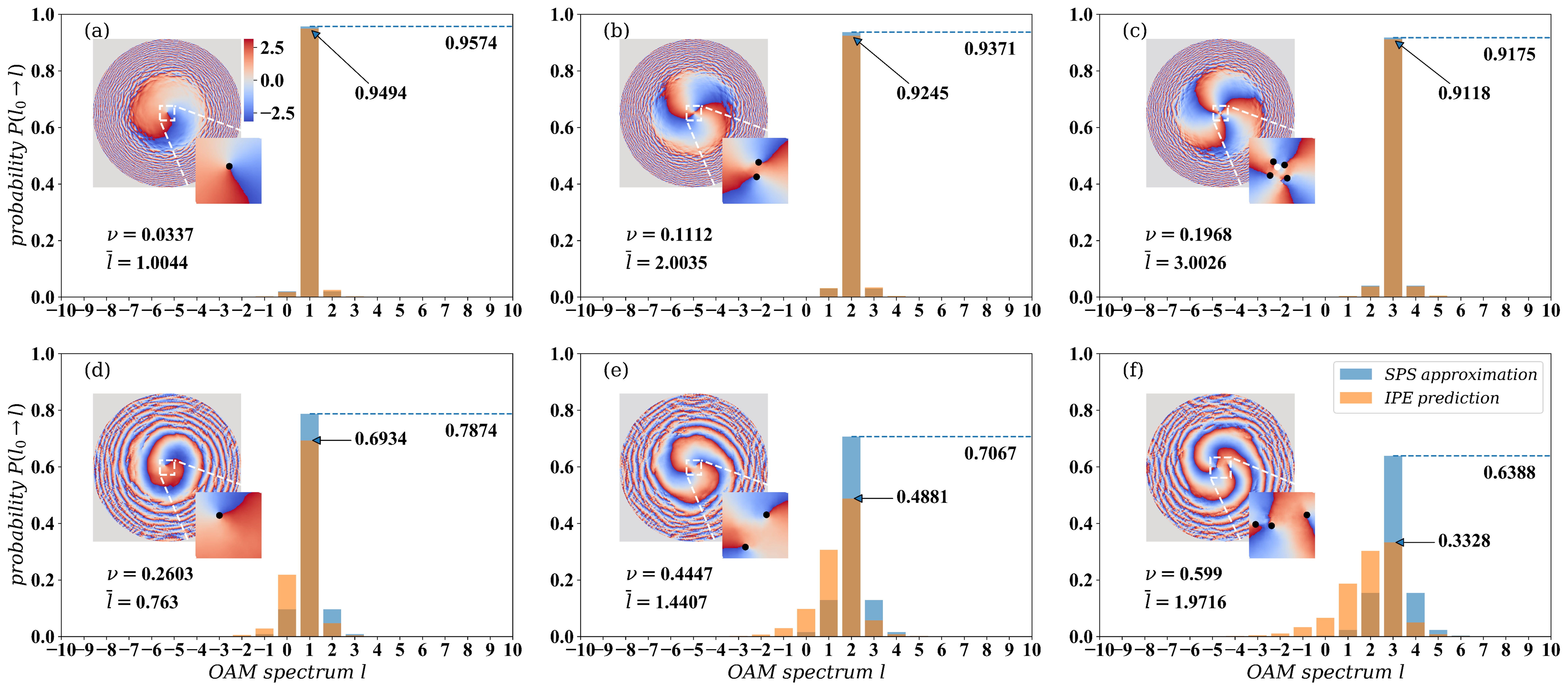}
\caption{OAM spectra averaged over 300 realizations of turbulence and phase
distributions of the received optical field for a single realization of
turbulence with different azimuthal indices: (a), (d) $l_{0}=1$, (b), (e) $%
l_{0}=2$, (c), (f) $l_{0}=3$. The top row represents the results below the
splitting-thresholds ((a) $\protect\nu =0.0337$; (b) $\protect\nu =0.1112$;
(c) $\protect\nu =0.1968$), corresponding to $t=0.157$ approximately. The
bottom row represents the results above the splitting-thresholds ((d) $%
\protect\nu =0.2603$; (e) $\protect\nu =0.4447$; (f) $\protect\nu =0.599$),
corresponding to $t=1.376$ approximately. The colorbar for all phase
distributions (a)--(f) are the same. The dashed lines denote the
eigen-crosstalk probabilities obtained from the results of the SPS
approximation and the arrows point to the eigen-crosstalk probabilities
obtained from the results of the IPE prediction. The bi-color dots (i.e.,
black dots and white dots) plotted in each partial enlarged plot indicate
the location of individual vortices, corresponding to a TC of $+1$ and a TC
of $-1$, respectively.}
\label{fig:6}
\end{figure*}

On the other hand, we also verify our anticipation in Fig. \ref{fig:6} with
a more straightforward way, which is a evident illustration of the phase
distributions of the received optical field. The OAM spectra of the received
optical field, averaged over 300 realizations of turbulence under the weak
scintillation condition, are also plotted in Fig. \ref{fig:6}. The top row
in Fig. \ref{fig:6} represents the results below the splitting-threshold
(e.g., (a) $\nu =0.0337$, (b) $\nu =0.1112$, (c) $\nu =0.1968$,
corresponding to $t=0.157$ approximately), while the bottom row represents
the results above the splitting-threshold (e.g., (d) $\nu =0.2603$, (e) $\nu
=0.4447$, (f) $\nu =0.599$, corresponding to $t=1.376$ approximately). By
comparing the results between top and bottom, we also obtain the same
conclusions presented in Fig. \ref{fig:5}, which is exactly the kind of
result we expected. More importantly, we observe in the top row of Fig. \ref%
{fig:6} that a pair of vortex-antivortex pairs occurs in the phase
distribution of the received optical field even when the vortex-splitting
ratio is below the splitting-threshold (see Fig. \ref{fig:6}(c)), which
makes us more confident to the above physical interpretation of the
deviation scale.

At last, we conclude that the appearance of deviation scale cannot be
predicted only by the Rytov variance, which can be predicted through the
vortex-splitting ratio of the received optical field alone or with the help
of $t$. In other words, it is not completely scientific for us to employ the
SPS approximation to simulate the crosstalk evolution of atmospheric
propagation even when the turbulent channel is under the weak scintillation
condition.

\section{CONCLUSION}
\label{sec:5}
In this paper, we give a suitable physical interpretation for
the recently so-called deviation scale [C. M. Mabena et al., Phys. Rev. A
99, 013828 (2019)], which bridges the connection between the result of the
IPE prediction and that of the SPS approximation. Before we explain the
nature of the deviation scale, we reexamine whether the minimal set of
parameters obtained by the IPE remains valid to completely determine the
crosstalk evolution considering the radial-mode scrambling. The process of
re-examination extends the results of the IPE and proofs the validity of our
numerical simulation. In our endeavor to present a qualitative understanding
of the deviation scale, we found that when the scintillation condition,
measured by the Rytov variance, is weak, namely, $\sigma _{R}^{2}<1$ and the
normalized propagation distance is below than a specific threshold, namely, $%
t<0.5$, the evolution curve of eigen-crosstalk probability almost coincides
with that of the SPS approximation. However, when $t>0.5$, the deviation
scale starts to appear and becomes larger as $\sigma _{R}^{2}$ increases,
which qualitatively gives a rough application scope of the SPS approximation
for an OAM-carrying beam propagating through atmospheric turbulence.
Furthermore, we demonstrate what the deviation scale actually represents
from some quantitative calculation. Our quantitative understanding is that
the spatial accumulation of slight intensity modulation of the incident
OAM-carrying beam splits the original vortex into multiple individual
vortices with a TC of $+1$ and re-generates the vortex-antivortex pairs with
a TC of $+1$ and with a TC of $-1$, leading to a significant reduction of
the value of the average OAM in the output plane and the deviation between
the results of the IPE prediction and that of the SPS approximation.
Notably, such phenomena will happen only if the disruption of this compound
effect on the phase distribution of the incident OAM-carrying beam becomes
more significant, which can be quantified by vortex-splitting ratio of the
received optical field. Or to put in another way, such phenomena will happen
only if the vortex-splitting ratio is beyond the splitting-threshold (e.g., $%
\nu =0.1$ for $l_{0}=1$, $\nu =0.15$ for $l_{0}=2$ and $\nu =0.2$ for $%
l_{0}=3$). In fact, the aforementioned two interpretations are equivalent
because the qualitative description gives a complete parameter condition
(i.e., $\sigma _{R}^{2}$ and $t$) composed of beam parameters and turbulence
parameters (i.e., $C_{n}^{2}$, $z$, $w_{0}$ and $\lambda $) to determine the
appearance of the deviation scale, while the quantitative one is obtained
from the splitting-threshold condition that completely and comprehensively
evaluates the results of the compound effect in the output plane, which can
be easily translated into the threshold of $t$ with different values (e.g., $%
t=0.648$ for $l_{0}=1$; $t=0.369$ for $l_{0}=2$; $t=0.255$ for $l_{0}=3$).
Generally, the quantitative calculation offers a more precise application
scope of the SPS approximation. On the other hand, these conclusions also
reveal that the appearance of deviation scale cannot be predicted only by
the Rytov variance, which can be measured with the help of $t$ or predicted
through the vortex-splitting ratio of the received optical field alone.\\

\noindent\textbf{Funding.} Anhui Provincial Natural Science Foundation
(1908085QA37); National Natural Science Foundation of China (11904369);
State Key Laboratory of Pulsed Power Laser Technology Supported by Open
Research Fund of State Key Laboratory of Pulsed Power Laser Technology
(2019ZR07).\\

\noindent\textbf{Acknowledgment.} We would like to thank the anonymous
reviewers for their valuable comments, which significantly improves the
presentation of this paper. We acknowledge the helpful discussion with
Professor Filippus Stef Roux. Zhiwei Tao thanks Professor Ruizhong Rao for
his insightful suggestions.\\

\noindent\textbf{Disclosures.} The authors declare no conflicts of interest.


\ifthenelse{\equal{\journalref}{aop}}{%
\section*{Author Biographies}
\begingroup
\setlength\intextsep{0pt}
\begin{minipage}[t][6.3cm][t]{1.0\textwidth} 
  \begin{wrapfigure}{L}{0.25\textwidth}
    \includegraphics[width=0.25\textwidth]{john_smith.eps}
  \end{wrapfigure}
  \noindent
  {\bfseries John Smith} received his BSc (Mathematics) in 2000 from The University of Maryland. His research interests include lasers and optics.
\end{minipage}
\begin{minipage}{1.0\textwidth}
  \begin{wrapfigure}{L}{0.25\textwidth}
    \includegraphics[width=0.25\textwidth]{alice_smith.eps}
  \end{wrapfigure}
  \noindent
  {\bfseries Alice Smith} also received her BSc (Mathematics) in 2000 from The University of Maryland. Her research interests also include lasers and optics.
\end{minipage}
\endgroup
}{}


\begin{thebibliography}{99}
\bibitem {c29}It is worth noting that the link between TC and OAM is only
valid for rotationally symmetric beams, such as Laguerre-Gaussian modes and
Bessel-Gaussian modes, to name just a few. In all other cases, the OAM in the
beam is in general not proportional to the TC in the beam. The listed
rotationally symmetric beams carry the same OAM normalized to the optical
power, which is equal to the beams' integer TC. For the latest progresses
about TC, we refer the reader to Refs. \cite{c30,c31,c32,c33}.
\end{thebibliography}
\end{document}